# Electrical and Magnetic behaviour of $PrFeAsO_{0.8}F_{0.2}$ superconductor

R S Meena[1,2], Anand Pal[1], K V R Rao[2], Hari Kishan[1] and V.P.S. Awana[1]*

[1]Quantum Phenomena and Application Division, National Physical Laboratory (CSIR)

Dr. K. S. Krishnan Road, New Delhi-110012, India

[2]Department of Physics, University of Rajasthan, Jaipur-302055, India

The superconducting and ground state samples of $PrFeAsO_{0.8}F_{0.2}$ and $PrFeAsO$ have been synthesised *via easy and versatile single step solid state reaction route*. X-ray & Reitveld refine parameters of the synthesised samples are in good agreement to the earlier reported value of the structure. The ground state of the pristine compound (PrFeAsO) exhibited a metallic like step in resistivity below 150K followed by another step at 12K. The former is associated with the spin density wave (SDW) like ordering of Fe spins and later to the anomalous magnetic ordering for Pr moments. Both the resistivity anomalies are absent in case of superconducting $PrFeAsO_{0.8}F_{0.2}$ sample. Detailed high field (up to 12Tesla) electrical and magnetization measurements are carried out for superconducting $PrFeAsO_{0.8}F_{0.2}$ sample. The $PrFeAsO_{0.8}F_{0.2}$ exhibited superconducting onset ($T_c^{onset}$) at around 47K with $T_c(\varrho=0)$ at 38K. Though the $T_c^{onset}$ remains nearly invariant, the $T_c(\varrho=0)$ is decreased with applied field, and the same is around 23K under applied field of 12Tesla. The upper critical field ($H_{c2}$) is estimated from the Ginzburg Landau equation (GL) fitting, which is found to be ~ 182Tesla. Critical current density (Jc) being calculated from high field isothermal magnetization (MH) loops with the help of Beans critical state model, is found to be of the order of $10^3$ A/cm$^2$. Summarily, the superconductivity characterization of single step synthesised $PrFeAsO_{0.8}F_{0.2}$ superconductor is presented.

**Introduction:**

Interest of superconductivity in scientific community rejuvenated with the happening of a breakthrough in Feb 2008, when superconductivity at 26K was realised in F doped

LaFeAsO compound by a Japanese group led by Prof. Hosono [1]. With great efforts of experimentalists with in a short span six different families of iron based superconductors were realised [2]. All these families have a common structural feature i.e., existence of a tetrahedral Fe-pnictogen (As/P) or Fe-chalcogen(Se/Te) layer in which iron atoms are sitting on the corner of square plane and pnictogen/chalcogen sitting above and below to this plane. This tetrahedral layer is separated from the rare-earth and oxygen/fluorine layer. The Fe-chalcogen/Pnictogen planes are the main building blocks responsible for appearance of superconductivity in these iron based superconductors, which are sandwiched by other layers. Other layers on the other hand play a vital role in determining the electronic properties of superconducting block either by donating charges or creating internal pressure to the FeAs/FeSe layers. Interestingly these quaternary oxy-pnictides type compounds have been synthesised way back by a German group with ZrCuSiAs type structure and a general formula REOMPn, where RE= La, Ce, Pr, Nd, Sm, Eu and Gd, etc, M =Mn, Fe, Co and Ni, etc and Pn= P and As [3]. Some of these compounds show superconductivity at low temperature, forming a new family of superconductors other than cuprates with a layered structure [4]. Pristine REFeAsO (RE=rare earth) compounds are non superconducting and do exhibit a structural change from tetragonal to orthorhombic at around 150K followed by an AFM like ordering of Fe spins and finally RE magnetic ordering at low temperature [5-11]. The REFeAsO show superconductivity of up to 55K (RE=Sm) by $O^{-2}$ site $F^{-1}$ doping [12-13]. Theoreticians studied the role of electron doping and point out that 3d orbitals of Fe give the Fermi surface a hole at $\Gamma$-point and an electron at M-point in a reduced Brilliouin zone [14, 15]. Spin density wave state (SDW) or anti-ferromagnetic ordering is due to nesting of the $\Gamma$-point and M-point in the un-doped pristine REFeAsO compounds. Electron doping suppresses SDW and superconductivity appears because of the dynamical spin fluctuations caused by nesting [16-20].

At low temperature magnetic moments on the rare-earth sites are ordered anti-ferro-magnetically, which is obviously missing in case of nonmagnetic La [5-9]. Pr in PrFeAsO showed highest $T_N^{Pr}$ = 12K [5], which is attributed to sensitive interaction between Fe and Pr moments as revealed by Mossbauer [21] and μSR [22] studies. The anomalous ordering of Pr ($T_N^{Pr}$ = 12 K) in PrFeAsO has been studied in detail by specific heat (Cp) and negative thermal expansion (NTE) experiments [23-25]. Keeping in view the anomalous ordering of Pr in PrFeAsO [5, 21-25], here we study the high field electrical and magnetic properties of the pure and fluorine doped PrFeAsO and PrFeAsO$_{0.8}$F$_{0.2}$ samples, which are prepared by the single step method.

**Experimental:**

We took Pr ingots and powder of As, PrF$_3$, Fe, and Fe$_2$O$_3$ in their stoichiometry ratio to prepare the polycrystalline samples with nominal composition PrFeAsO and PrFeAsO$_{0.8}$F$_{0.2}$. The ingots and the powders are weighed and grinded in a controlled atmosphere of Glove Box. The grinded sample is palletized and sealed in an evacuated quartz tube prior to sintering. The sample is sintered at 550C for 8 hours, followed by 950C for 12 hours and finally at 110 C for 33 hours. The detailed procedure of this single step preparation had been reported by us in ref. [26]. The phase purity and the room temperature structural parameters were determined by powder X-ray diffraction pattern using CuK$_\alpha$ radiation on Desktop Rigaku-Miniflex II diffractometer. Reitveld analysis is performed with the help of Fullprof programme. The resistivity and magnetization measurements were carried on a Quantum Design Physical Property (PPMS) measurement system.

**Results and Discussion:**

The room temperature XRD pattern for PrFeAsO and PrFeAsO$_{0.8}$F$_{0.2}$ samples along with their Rietveld refinements are shown in Figure 1. It is clear from the Figure that all main peaks are well indexed on the basis of tetragonal ZrCuSAs structure with space group P4/nmm. All the permitted diffraction planes of these samples are marked with blue vertical lines between the observed and fitted patterns and their difference by the green lines, in the bottom. Minor impurity peaks are also seen and marked with *, which are mainly from un-reacted residuals in fluorine doped sample. The lattice parameters obtained are a = 3.981(4) Å and c = 8.593(1)Å for PrFeAsO sample and a = 3.972(3)Å and c = 8.572(3)Å for PrFeAsO$_{0.8}$F$_{0.2}$. Shrinking of lattice parameters is witnessed on the partial substitution of fluorine at oxygen site. These values are comparable with the earlier reported values [5, 21-25]. The lattice parameters of Pr samples are between Ce and Nd counter parts of REFeAsO/F series [27-29]. The standard Wyckoff positions being used for Rietveld refinements of PrFeAsO are shown in Table 1. The typical Rietveld refined parameters including lattice constants, volume, the quality of fit ($\chi^2$) and fitted Z$_{Pr}$/Z$_{As}$ values for PrFeAsO and PrFeAsO$_{0.8}$F$_{0.2}$ are given in Table 2.

Figure 2 (a) shows the resistivity versus temperature plots for the PrFeAsO and PrFeAsO$_{0.8}$F$_{0.2}$ samples at zero applied magnetic fields. The ground state PrFeAsO is not superconducting down to 2K, rather it shows a steep metallic step at around 150K. The same is attached to structural transformation accompanied by a spin density wave (SDW) transition of Fe moments for various REFeAsO (RE=La, Ce, Nd, and Sm) compounds [30-31]. The inset of Fig 2(a) depicts the derivative of resistivity with temperature (d$\varrho$/dt). A distinctive peak is seen at 12K in d$\varrho$/dT vs T plot, which is assigned to anti-ferromagnetic (AFM) ordering of Pr moments (T$_N^{Pr}$) [5, 21-25]. The origin of the anomalous ordering is due to interacting Pr and Fe moments, as confirmed by neutron diffraction experiments [29, 31]. The magnetic ordering of rare earth elements appear at low temperature below T$_N^{RE}$, which is 4.4K for Ce, 4.1K for Nd and 5.4K for Sm [7, 24, 31]. PrFeAsO has the highest RE ordering temperature of around 12K. The resistivity of superconducting PrFeAsO$_{0.8}$F$_{0.2}$ sample decreases with temperature, and an onset of superconducting transition occurred at around 47K followed by superconductivity with T$_c$($\varrho$=0) at 38K. The transition is rather broad accompanied with a long tail near $\varrho$=0 state. C.R. Rotundu et al, ref. [32] investigated detailed phase diagram of PrFeAsO$_{1-x}$F$_x$. The structural and anti-ferromagnetism phase suppression starts with 8 % Fluorine doping, and optimum superconductivity is achieved at 15 % to 22 % doping of fluorine. The T$_c$ of PrFeAsO$_{0.8}$F$_{0.2}$, in our case, is slightly lower as compared to others [28,32], could be due to slightly lesser amount of fluorine in the sample, which is difficult to identify exactly. Ionic size of rare-earth also plays pivotal role in deciding the transition temperature (T$_c$) [27-28], in the case of NdFeAsO$_{0.8}$F$_{0.2}$ and SmFeAsO$_{0.8}$F$_{0.2}$ the T$_c$ is found to be 48K and 53K respectively [33, 34]. Figure 2 (b) shows the temperature dependence of electrical resistivity with respect to varying magnetic field i.e., $\varrho$(T)Vs H under 0.1 to 12Tesla applied magnetic field. The resistance drops to almost zero at ~ 38 K in zero field. Under applied magnetic field though the T$_c^{onset}$ remains nearly invariant, the T$_c$($\varrho$ =0) is decreased to lower temperatures. This means the superconducting transition width becomes wider with applied field. Here we define upper critical field H$_{c2}$ as transition temperature T$_c$(H) which is equal to 90% of the normal state value ($\varrho_N$) for the applied field H. The T$_c$(H) values are 46.16K and 43.91 K at 0 and 12Tesla respectively. The calculated value of dH$_{c2}$/dT is around 4Tesla/K. The transition width $\Delta T_c = T_c(90\%) - T_c(10\%)$ are 7.64K and 14.7K for 0 and 12 Tesla respectively. The broadening of the between T$_c^{onset}$ and T$_c$($\varrho$=0) with applied field is considered to be due to the weak links between grains and magnetic flux flow through the vortices [8]. Figure 2(c) shows the temperature dependence of $H_{c2}$ for 90%, 50% and 10 % of $\varrho_N$. In all cases obvious sharp increase in $H_{c2}$ is noticed with decreasing temperature along with a weak upward curvature near transition ($T_c$) region. Similar observations have been reported in other ReFeAsO$_{1-x}$F$_x$ compounds [34-36]. Figure 2(d) shows the variation of upper critical field, which is determined by using different criteria of $H_{c2}$ at $\varrho$=90%, 50% and 10% of $\varrho_N$, here $\varrho_N$ is normal resistivity. The $H_{c2}$ variation with temperature is depicted in figure 2(d), we used Ginzburg Landau (GL) equation to find the $H_{c2}(0)$ value. $H_{c2}(T) = H_{c2}(0)(1-t^2)/1+t^2$, where t =$T/T_c$ is the reduced temperature [37]. The estimated values for our sample at $H_{c2}(90\%)$, $H_{c2}(50\%)$ and $H_{c2}(10\%)$ are 182Tesla, 73Tesla and 41.6Tesla respectively.

Figure 3(a). Shows the iso-thermal magnetization M(H) plots for PrFeAsO$_{0.8}$F$_{0.2}$ at 5, 10, 20 and 30K. The lower critical field ($H_{c1}$) is determined as the deviation from the linear MH behaviour of the curve. The lower critical field ($H_{c1}$) values are observed at round 120, 220 and 340Oe at 30, 20 and 10K respectively. Figure 3(b) Show the iso-thermal magnetization loops of the studied compound PrFeAsO$_{0.8}$F$_{0.2}$ at 5 and 50K, in higher applied fields of up to 7kOe. The critical current density (Jc) of the PrFeAsO$_{0.8}$F$_{0.2}$ sample has been estimated by using Beans critical state model. For a bar shape sample, the magnetic critical current density can be calculated using the relation Jc = 20ΔM/a(1-a/3b), here ΔM is the width of hysteresis loop and a and b are dimensions of the sample, with a<b and the magnetic field direction is perpendicular to the plane. The critical current density roughly estimated from the observed M(H) curve is order of ~$10^3$A/cm$^2$ at 5K. The critical current density of a polycrystalline samples are far less than the estimated values of the single crystal. The critical current density for a single crystal of PrFeAsO$_{0.7}$ ($T_c$=35K), is reported 2.9 x $10^5$A/cm$^2$ by Kashiwaya et al. [38]. Figure 4 shows the DC magnetic susceptibility ($\chi$) in zero field cooled (ZFC) and field cooled (FC) situations in the temperature range 5 to 50K under an applied field of 10 Oe. At around 42K a diamagnetic onset transition appears which shows bulk nature of superconductivity in studied PrFeAsO$_{0.8}$F$_{0.2}$ sample. The estimated value of volume fraction ~ 7% in meissner state and the shielding fraction is around 16%. Because of the impurities, pinning defects and the magnetic contribution of the Pr/Fe ions determination of exact value of volume fraction is not possible.

In conclusion, we have synthesised pristine and superconducting sample of PrFeAsO with fluorine doping by an easy single step method. The ground state of the pristine compound (PrFeAsO) exhibited magnetic ordering of Fe moments at around 150K, followed by the unusual anti-ferromagnetic ordering of Pr at 12K. Both these transitions are missing in case of F doped superconducting $PrFeAsO_{0.8}F_{0.2}$ sample. The $PrFeAsO_{0.8}F_{0.2}$ sample exhibited bulk at below 42K. The sample is characterised for its magnetic and electrical behaviour under 12Tesla field. The upper critical field is estimated to be ~182Tesla. The critical current density ($J_c$) value as calculated by applying Bean's critical model which is of order of ~$10^3$ A/cm$^2$.

Authors would like to thank Prof. R C Budhani, Director, NPL for his keen interest and encouragement for carrying out the present work. A Pal would like to thank DAE-SRC outstanding award scheme for financial support as post doctoral fellow. H Kishan thanks CSIR for providing Emeritus Scientist fellowship.

**References:**

1. Y. Kamihara, T. Watanabe, M. Hirano, H. Hosono; J. Am. Chem. Soc. **130**, 3296, (2008)
2. H. H. Wen and S Li; Annu. Rev Condensed. Matter Physics **2,** 211 (2011)
3. B. I. Zimmer, W. Jeitschko, J. H. Albering, R. Glaum, M. Reehuis; J. Alloys Comp. **229**, 238 (1995)
4. Y. Kamihara, H. Hiramatsu, M. Hirano, R. Kawamura, H. Yanagi, T. Kamiya, H. Hosono; J. Am. Chem. Soc. **128**, 10 012 (2006)
5. J. Zhao, Q. Huang, C. de la Cruz, J. W. Lynn, M. D. Lumsden, Z. A. Ren, J. Yang, X. Shen, X. Dong, Z. Zhao, and P. Dai; Phys. Rev. B **78**, 132504 (2008)
6. Y. Chen, J.W. Lynn, J. Li, G. Li, G. F. Chen, J. L. Luo, N. L.Wang, P. Dai, C. dela Cruz, and H. A. Mook; Phys. Rev. B **78**, 064515 (2008)
7. L. Ding, C. He, J. K. Dong, T. Wu, R. H. Liu, X. H. Chen, and S. Y. Li; Phys. Rev. B **77**, 180510(R) (2008)
8. Y. Qiu, Wei Bao, Q. Huang, T. Yildirim, J. M. Simmons, M. A. Green, J. W. Lynn, Y. C. Gasparovic, J. Li, T. Wu, G. Wu, and X. H. Chen; Phys. Rev. Lett. **101**, 257002 (2008)
9. C. Wang, L. Li, S. Chi, Z. Zhu, Z. Ren, Y. Li, Y. Wang, X. Lin, Y. Luo, S. Jiang, X. Xu, G. Cao, and Z. Xu; Europhys. Lett. **83** 67006 (2008)
10. P. Wang, Z. M. Stadnik, C. Wang, G.-H. Cao, and Z.-A. Xu; J. Phys. Condens. Matter **22**, 145701 (2010)
11. A. Alfonsov, F. Mur´anyi, V. Kataev, G. Lang, N. Leps, L. Wang, R. Klingeler, A. Kondrat, C. Hess, S. Wurmehl, A. K¨ohler, G. Behr, S. Hampel, M. Deutschmann, S.

Katrych, N. D. Zhigadlo, Z. Bukowski, J. Karpinski, and B.Büchner; Phys. Rev. B **83**, 094526 (2011)

12. Ren ZA, Lu W, Yang J, Yi W, Shen XL, et al.; Chin. Phys. Lett. **25,** 2215 (2008)
13. Wang C, Li L, Chi S, Zhu Z, Ren Z, et al.; Europhys. Lett. **83,** 67006 (2008)
14. D. J Singh and M.H.Du; Phys. Rev. Lett. **100,** 237003 (2008)
15. Katsuaki Kodama, Motoyuki Ishikado, Fumitaka Esaka, Akira Iyo, Hiroshi Eisaki, Shin-ichi Shamoto; Journal of the Physical Society of Japan vol. **80**, 034601 (2011)
16. I.I Mazin, D.J Singh, M.D Johannes, M.H Du; Phys. Rev. Lett. **101,** 057003 ( 2008)
17. V. Cvetkovic and Z. Tesanovic ; Europhys. Lett. **85**, 37002. (2009)
18. K. Kuroki, S. Onari, R. Arita, H. Usui, Y. Tanaka, H. Kontani, and H. Aoki ; Phys. Rev. Lett. **101,** 087004 (2008)
19. F. Ma and Z.-Y. Lu; Phys. Rev. B **78,** 033111 (2008)
20. S. Wakimoto, K. Kodama, M. Ishikado, M. Matsuda, R. Kajimoto, M. Arai, K. Kakurai, F. Esaka, A. Iyo, H. Kito, H. Eisaki, and S. Shamoto; J. Phys. Soc. Jpn. **79,** 074715 (2010)
21. M. A. McGuire, R. P. Hermann, A. S. Sefat, B. C. Sales, R. Jin, D. Mandrus, F. Grandjean, and G. J. Long; New J. Phys. **11**, 025011 (2009)
22. H. Maeter, H. Luetkens, Yu. G. Pashkevich, A. Kwadrin, R. Khasanov, A. Amato, A. A. Gusev, K. V. Lamonova, D. A. Chervinskii, R. Klingeler, C. Hess, G. Behr, B. Büchner, and H.-H. Klauss ;Phys. Rev. B **80**, 094524 (2009)
23. U. Stockert, N. Leps, L. Wang, G. Behr, S. Wurmehl, B. Büchner, and R. Klingeler ; Phys. Rev. B **86**, 144407 (2012)
24. S. A. J. Kimber, D. N. Argyriou, F. Yokaichiya, K. Habicht, S. Gerischer, T. Hansen, T. Chatterji, R. Klingeler, C. Hess, G. Behr, A. Kondrat, and B. Büchner; Phys. Rev. B **78**, 140503 (2008)
25. D.Bhoi P Mandal and P Choudhury;Supercond. Sci. Technol **21,** 125021 ( 2008)
26. R. S. Meena, S. K. Singh, A. Pal, A. Kumar, R. Jha, K. V. R. Rao, Y.Du, X. L. Wang, and V. P. S. Awana, J. Appl. Phys. **111**, 07E232 (2012)
27. Zhi-An Ren, Guang-Can Che, Xiao-Li Dong, Jie Yang, Wei Lu, Wei Yi, Xiao-Li Shen, Zheng-Cai Li, Li-Ling Sun, Fang Zhou and Zhong-Xian Zhao; Eur. Phys. Let. **83,** 17002 ( 2008)
28. Z. A. Ren, J. Yang, W. Lu, W. Yi, G. C. Che, X. L. Dong, L. L. Sun and Z. X. Zhao; Mater. Res. Innovations **12**, 105 (2008)

29. Michael A McGuire, Rapha P Hermann, Athena S Sefat, Brian C Sales, Rongying Jin, David Mandrus, Fernande Grandjean and Gary J Long; New Journal of Physics **11,** 025011 (2009)
30. R. H. Liu, G. Wu, T. Wu, D. F. Fang, H. Chen, S. Y. Li, K. Liu, Y. L. Xie, X. F. Wang, R. L. Yang, L. Ding, C. He, D. L. Feng, and X. H. Chen; Phys. Rev. Lett. **101,** 087001 (2008)
31. J. Zhao, Q. Huang, C. de la Cruz, J. W. Lynn, M. D. Lumsden, Z. A. Ren, J. Yang, X. Shen, X. Dong, Z. Zhao, and P. Dai; Phys. Rev. B **78,** 132504 (2008).
32. C. R. Rotundu, D. T. Keane, B. Freelon, S. D. Wilson, A. Kim, P. N. Valdivia, E. Bourret-Courchesne, R. J. Birgeneau; arXiv:0907.1308v1
33. V.P.S. Awana, R.S. Meena, A. Pal, A. Vajpayee, K.V.R. Rao, and H. Kishan; Eur. Phys. J. B **79,** 139–146 (2011)
34. R.S. Meena, Anand Pal, S. Kumar, K.V. R. Rao and V P S Awana; J Supercond Non Magn. **26,** 2383 ( 2013)
35. D. Bhoi, L S Sharath Chandra, P Choudhary, V Ganeshan and P Mandal; Superconductor Sci. Technol **22**, 095015 (2009)
36. J. Jaroszynski, Scott C. Riggs, F. Hunte et.al; Phys. Rev. B **78,** 064511 (2008)
37. A Gurevich; Phys.Rev B **67,** 184515 (2003)
38. H. Kashiwaya, K. Shirai, T. Matsumoto, H. Shibata, H. Kambara, M. Ishikado, H. Eisaki, A. Iyo, S. Shamoto, I. Kurosawa, and S. Kashiwaya; Applied Physics Letters **96,** 202504 (2010)
39. M. Nikolo R.B. Goldfarb; Phys. Rev. B **39,** 6615 (1989)
40. G. Bonsignore, A. Agliolo Gallitto M. Li Vigni, J. L. Luo, G. F. Chen, N. L. Wang, D. V. Shovkun ; J. Low. Temp. Phys. **162,** 40-51 (2011)

**Captions:**

Figure 1: The room temperature X-ray diffraction patterns of PrFeAsO and PrFeAsO$_{0.8}$F$_{0.2}$ samples. The solid line represents the Rietveld refinement of the diffraction pattern with space group P4/nmm.

Figure 2(a): The temperature dependence of resistivity for the PrFeAsO and PrFeAsO$_{0.8}$F$_{0.2}$ samples synthesised by single step method, Inset, shows the dρ/dt versus T curve for PrFeAsO sample in extended low temperature regime.

Figure 2(b): Magneto-resistivity behaviour ρ(T) H for PrFeAsO$_{0.8}$F$_{0.2}$ sample under applied fields of up to 12Tesla.

Figure 2(c): Upper critical field H$_{c2}$ versus T for plots PrFeAsO$_{0.8}$F$_{0.2}$ superconductor at 90%, 50% and 10% normal resistivity criteria.

Figure 2 (d): Dependence of upper critical field H$_{c2}$(T) with temperature for PrFeAsO$_{0.8}$F$_{0.2}$ sample using Ginzburg-Landau (GL) equation for 90 %, 50 % and 10 % drop of the normal state resistivity.

Figure 3(a): Isothermal magnetization loops M(H) in the first quadrant at 5, 10, 20 and 30K for PrFeAsO$_{0.8}$F$_{0.2}$ superconducting sample.

Figure 3(b): Isothermal magnetization loop M(H) at 5 and 50K for superconducting PrFeAsO$_{0.8}$F$_{0.2}$ sample.

Figure 4: Zero field cooled (ZFC) and field cooled (FC) DC magnetic susceptibility plots for PrFeAsO$_{0.8}$F$_{0.2}$ sample.

Table 1 Wyckoff position for PrFeAsO (Space group: *P4/nmm*)

| Atom | Site | x | y | z |
|------|------|-----|-----|------|
| Pr | 2c | 1/4 | 1/4 | $Z_{Pr}$ |
| Fe | 2b | 3/4 | 1/4 | 1/2 |
| As | 2c | 1/4 | 1/4 | $Z_{As}$ |
| O/F | 2a | 3/4 | 3/4 | 0.00 |

Table 2 Reitveld refined parameters for PrFeAsO and PrFeAsO$_{0.8}$F$_{0.2}$

| Sample | $a^0$(A) | $c^0$(A) | Volume (A$^3$) | $Z_{Pr}$ | $Z_{As}$ | Rwp | $\chi^2$ |
|--------|----------|----------|----------------|----------|----------|------|------|
| PrFeAsO | 3.981(4) | 8.593(1) | 136.22 | 0.13755 | 0.65645 | 4.80 | 2.59 |
| PrFeAsO$_{0.8}$F$_{0.2}$ | 3.972(4) | 8.572(3) | 135.27 | 0.15556 | 0.65406 | 6.19 | 3.78 |

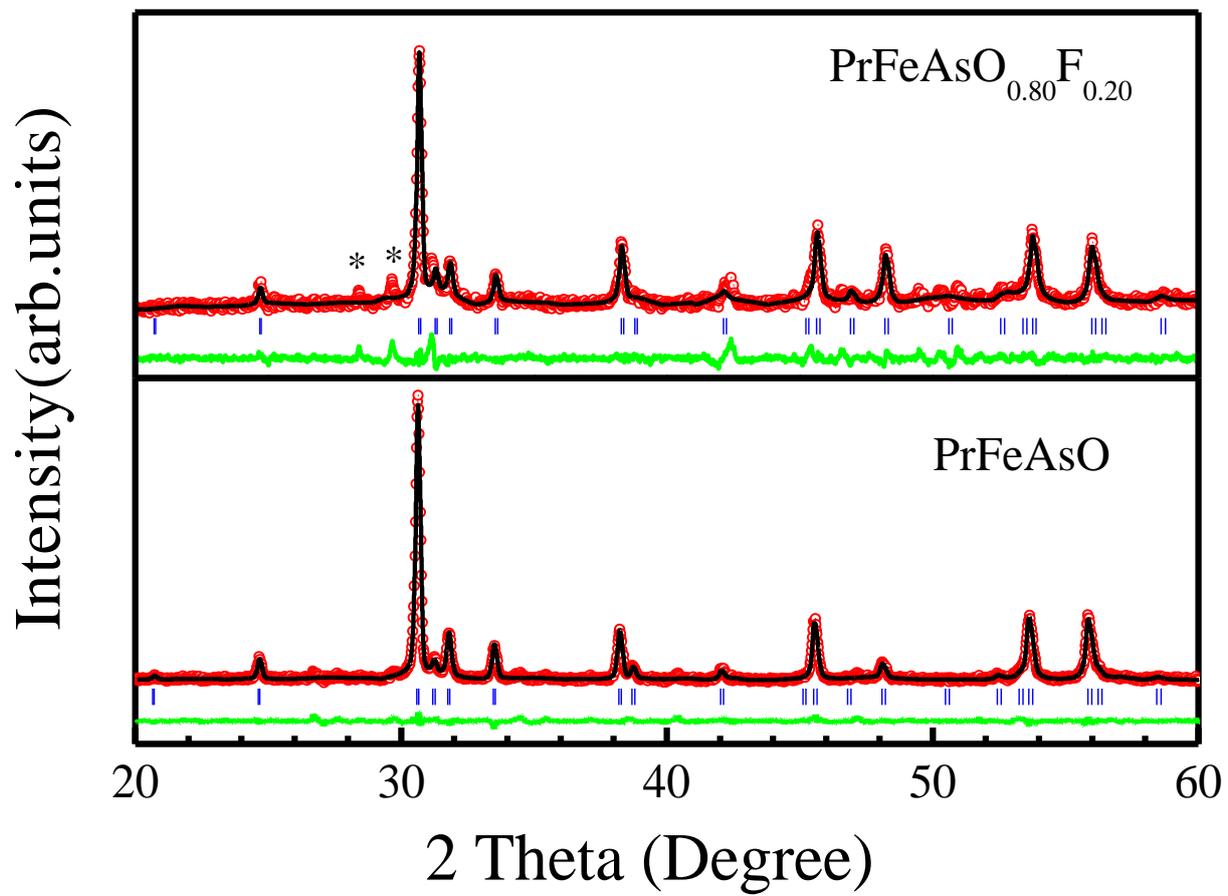

Figure 1

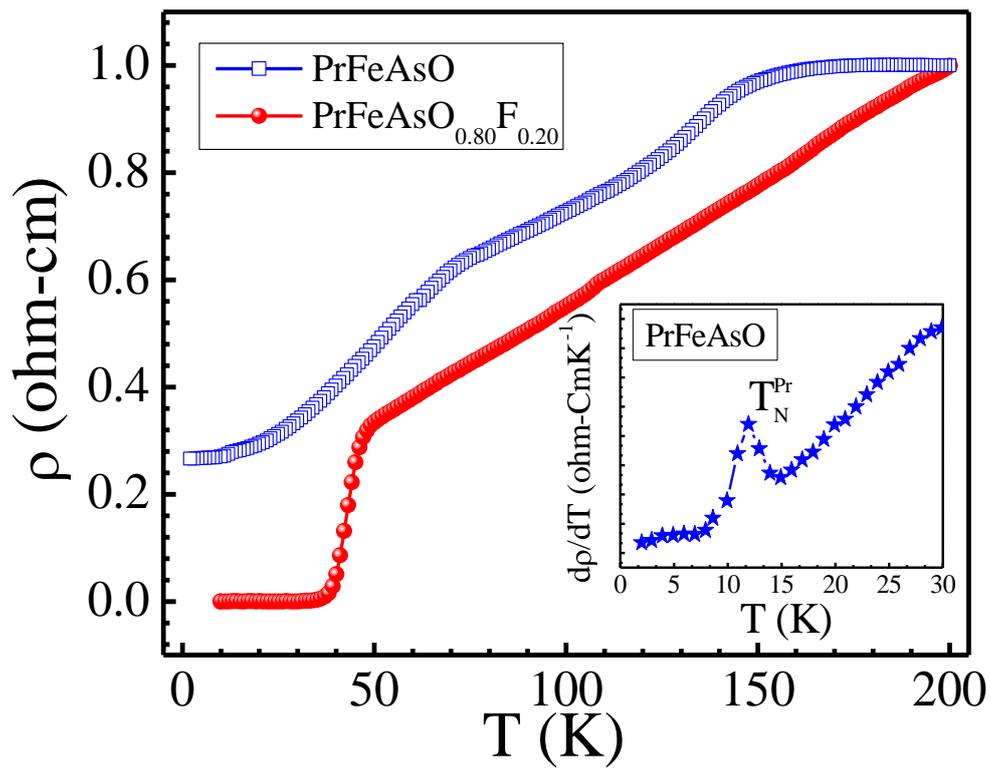

Figure 2(a)

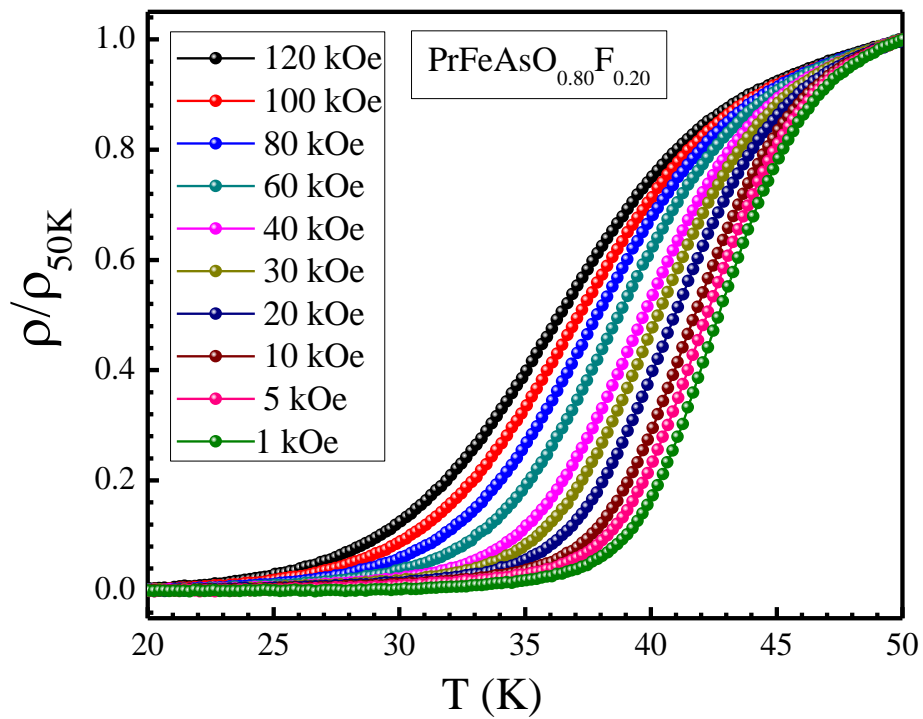

Figure 2(b)

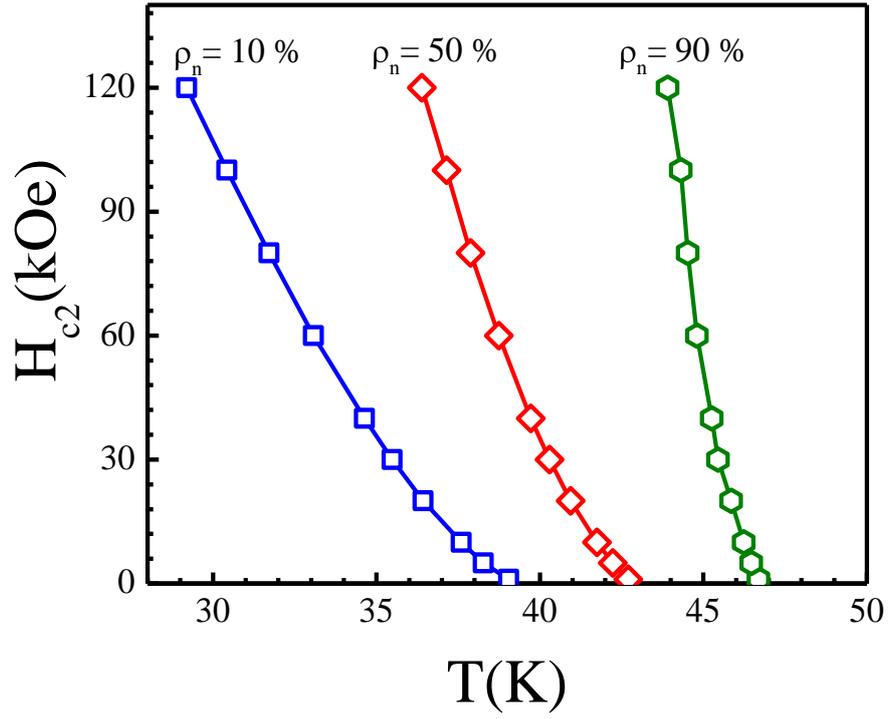

Figure 2(c)

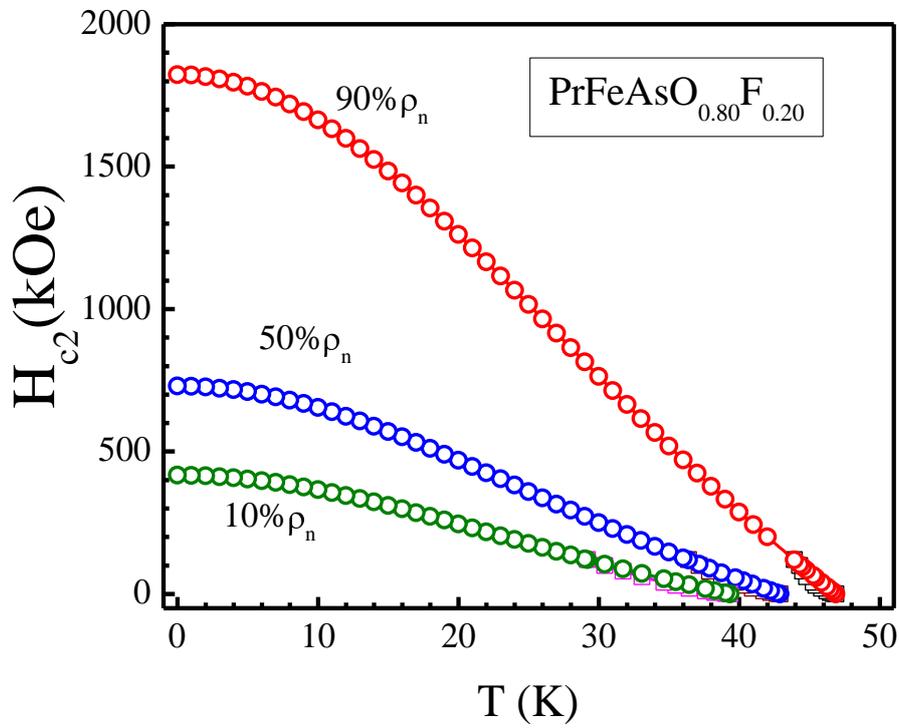

Figure 2(d)

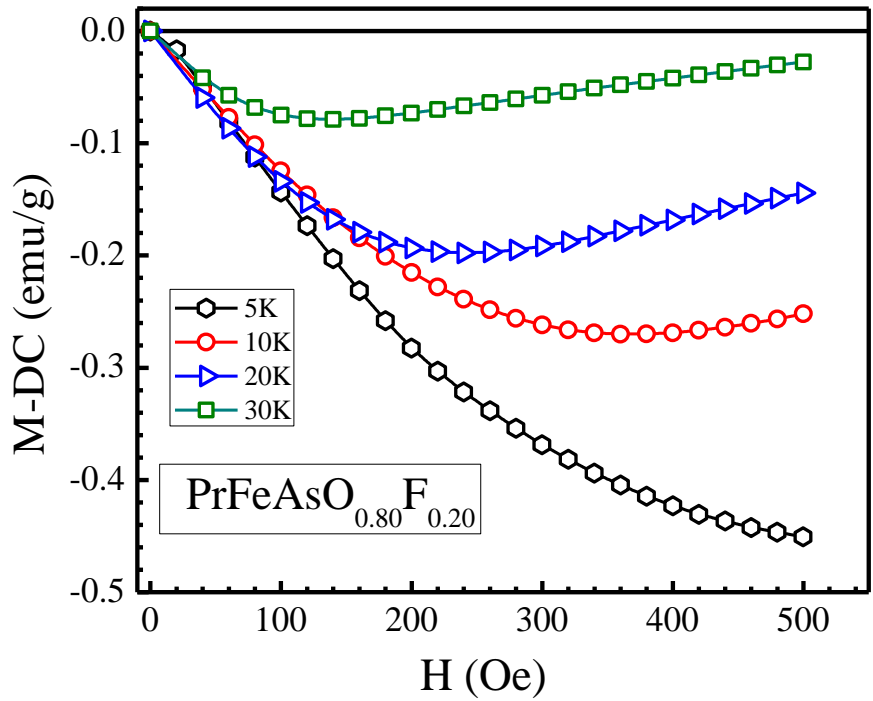

Figure 3(a)

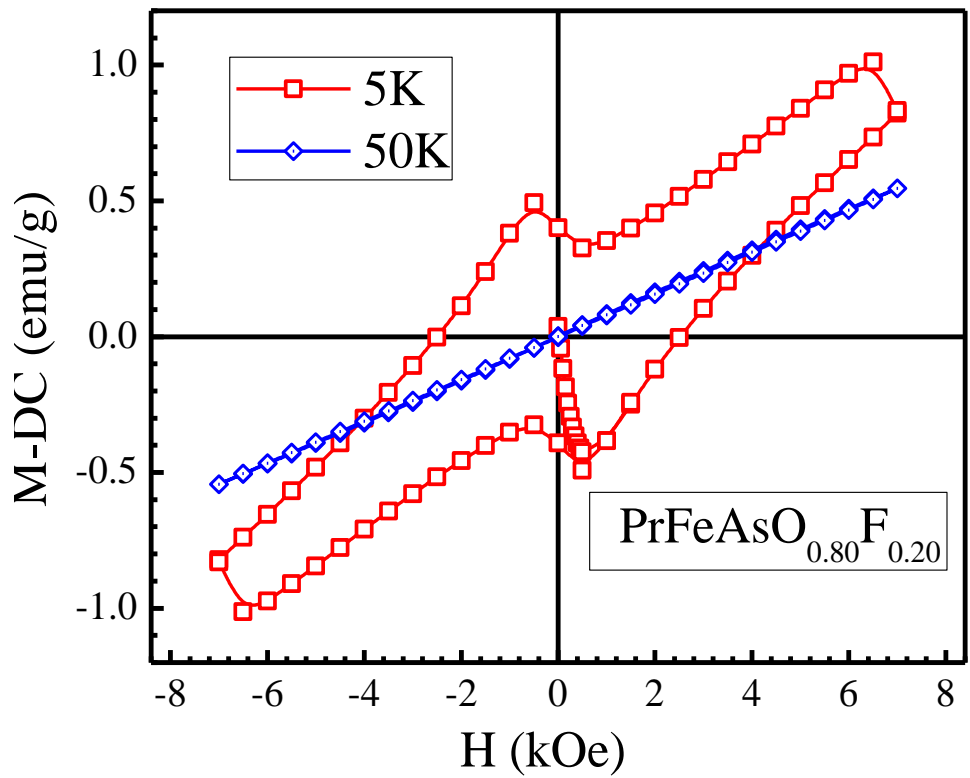

Figure 3(b)

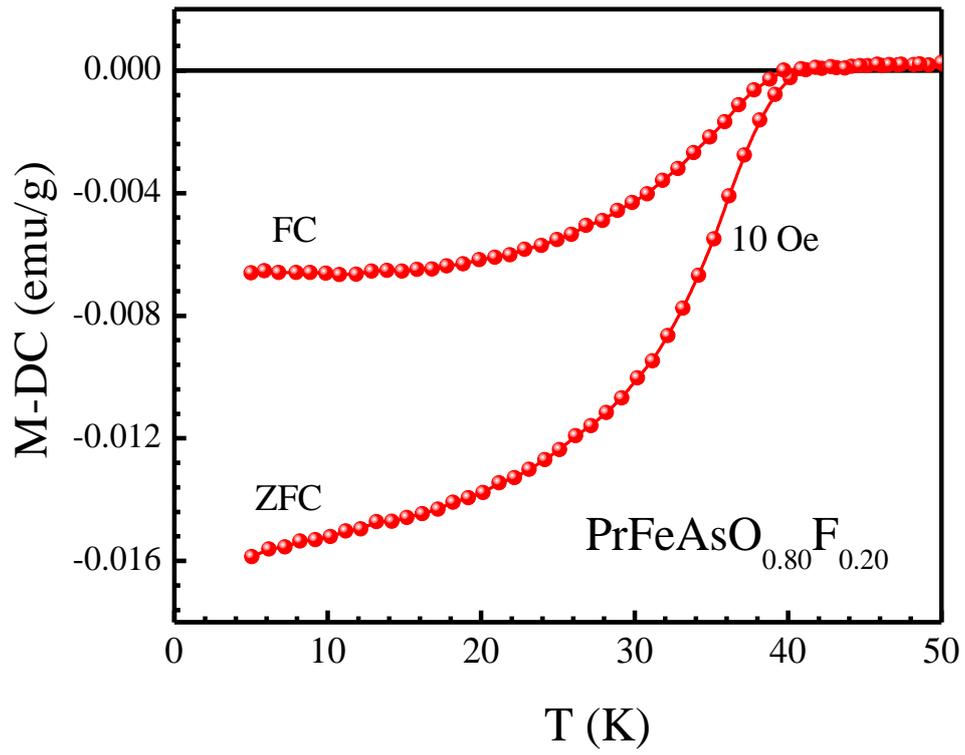

Figure 4